\documentstyle[aps,floats,epsf,twocolumn]{revtex}

\def\a{\alpha}
\def\b{\beta}
\def\d{\delta}
\def\D{\Delta}
\def\l{\lambda}
\def\w{\omega}
\def\g{\gamma}
\def\G{\Gamma}
\def\s{\sigma}
\def\t{\tau}

\def\be{\begin{equation}}
\def\ee{\end{equation}}
\def\ben{\begin{displaymath}}
\def\een{\end{displaymath}}
\def\bea{\begin{eqnarray}}
\def\eea{\end{eqnarray}}
\def\bean{\begin{eqnarray*}}
\def\eean{\end{eqnarray*}}

\def\swq #1#2{T\sum_{#1_n}\int \frac{d^2\vec#2}{(2\pi)^2}}
\def\iqe #1{\int \frac{d^3\vec#1}{(2\pi)^3}}
\def\iq #1#2{\int^{#1} \frac{d^3\vec#2}{(2\pi)^3}}
\def\itr #1#2{\int d#1 d^2\vec#2}
\def\su{\uparrow}
\def\sd{\downarrow}
\def\hc{\textrm{h.c.}}
\def\diag{\textrm{diag}}
\def\tr{\textrm{tr}}
\def\x{\otimes}
\def\pd #1{\partial_{#1}}
\def\I{\textbf{1}}
\def\av #1#2{\langle{#2}\rangle_{#1}}
\def\V{V}
\def\H{{\cal H}}

\begin{document}

\title{Fine structure of chiral symmetry breaking in the
       $QED_3$ theory of underdoped high-$T_c$ superconductors}

\author{Babak H. Seradjeh and Igor F. Herbut}

\address{Department of Physics, Simon Fraser University,
         Burnaby, British Columbia, \\
         Canada V5A 1S6\\}
\maketitle

\begin{abstract}
We study the effects of the leading irrelevant perturbations on chiral symmetry
breaking in the effective $QED_3$ theory of $d$-wave superconductor-insulator
transition in underdoped cuprates. For weak symmetry breaking, the effect of a
perturbation on energies of various insulating states can be classified according to
its engineering dimension in the maximally symmetric theory. Considering the velocity
anisotropy, repulsive interactions, and higher order derivatives we show that the
insulating state with the lowest energy is the spin density wave.
\end{abstract}


\section{Introduction}
The $d$-wave superconducting state, besides the familiar rotational invariance, at low
energies possesses an additional "chiral" symmetry for its quasiparticle
excitations~\cite{igor}. It was recently shown~\cite{igor} that the quantum phase
transition from the antiferromagnetic insulator to the $d$-wave superconductor may be
understood in terms of this hidden symmetry: while present in the superconducting
state, chiral symmetry is manifestly broken when the spin density wave (SDW) order
develops. Dynamical agent that brings about this change of symmetry was argued to be
provided by the vortex fluctuations of the condensate, that can be represented by an
effective gauge field minimally coupled to gapless quasiparticles of the $d$-wave
state~\cite{franz},~\cite{igor}. That way the theory of phase fluctuating $d$-wave
superconductor can be mapped onto the three dimensional quantum electrodynamics
($QED_3$), which was proposed as an effective low-energy description of the underdoped
high-temperature superconductors~\cite{igor}.

Chiral symmetry, however, is not an exact symmetry of the $d$-wave superconducting
state, but arises only in the low-energy limit of the standard quasiparticle action.
At low energies one is allowed to linearize the  spectrum near the nodes of the
$d$-wave order parameter and neglect the  higher order derivatives and the short-range
interactions between  quasiparticles, which are both linearly irrelevant by power
counting.  In this approximation the action becomes symmetric under a global
$U(2)\times U(2)$ transformation, where each $U(2)$ factor acts onto a four-component
Dirac field that describes the gapless excitations near the pair of the diagonally
opposed nodes \cite{igor} (see Fig.~\ref{nodes}). If one would  also neglect the
intrinsic velocity anisotropy of the $d$-wave superconductor  $v_F \gg v_\Delta$, with
$v_F$ being the Fermi velocity and  $v_\Delta$ the velocity related to the amplitude
of the superconducting order parameter, the chiral symmetry group would be enlarged
into the sixteen dimensional $U(4)$. The SDW insulator corresponds to breaking of this
symmetry along one particular "direction", while the broken generators of $U(4)$
rotate it towards one of the other possible broken symmetry states. Among these, one
can discern three additional types of insulators, related to the SDW (and to each
other) by chiral symmetry: the "$d+ip$" and "$d+is$" insulating states, and the
stripe-like one-dimensional charge density waves (CDW), in the classification of
refs.~\cite{igor} and~\cite{zlatko}.

In the case of the maximal $U(4)$ chiral symmetry, all the above insulating states,
and their various combinations \cite{igor}, are equally likely outcomes of the
spontaneous symmetry breaking.  This degeneracy is in reality removed  by the symmetry
breaking perturbations to the $QED_3$, most prominent of which have already been
mentioned above. For example, it was shown in~\cite{igor} that the short range
repulsion between quasiparticles favors the  SDW over the $d+ip$ insulator, and,
moreover, enhances the SDW order deeper in the insulating state. In this paper we
study the effects of weak velocity anisotropy, higher order derivatives, and
short-range repulsive interaction on the pattern of chiral symmetry breaking in the
$QED_3$ theory of the phase fluctuating $d$-wave superconductors. All of these
perturbations are irrelevant at low energies, and we show that their effects on the
energies of various states with weakly broken chiral symmetry  can be ordered
according to their engineering dimensionality:  velocity anisotropy, being only
marginally irrelevant~\cite{dom},~\cite{vafek}, provides then the dominant
perturbation. Weak repulsion and the second order derivatives are both equally
(linearly) irrelevant, but to the first order it is only the repulsion that affects
the energies of the insulating states. Fine structure of the chiral manifold of
insulators is schematically  depicted in Fig.~\ref{fine}.

Our main conclusion is that, to the leading order, it is the SDW state that is
selected by the weak perturbations to the $QED_3$. Although in real systems none of
the above perturbations is truly weak, we believe our result provides a useful
qualitative guide. In particular, it agrees with the standard picture of underdoped
cuprates, upon identification of the SDW insulator as being continuously connected
with the Mott insulating antiferromagnet near half filling.

In what follows we first briefly review the salient points of the $QED_3$ theory of
phase fluctuating $d$-wave superconductor, and set up the convenient notation in terms
of the eight component Dirac fields. We then proceed to calculate the lowest order
splittings of the energies due to velocity anisotropy, short-range repulsion, and
higher order derivative terms. We end with the summary and some concluding remarks on
our results.

\section{$QED_3$}
\begin{figure}[t]
  \epsfysize 70mm
  \epsfbox[-35 -15 300 320]{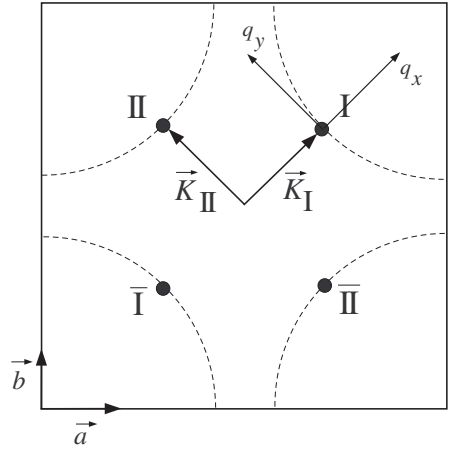}
  \noindent
  \caption[]{Four nodes of the $d$-wave order parameter. Dashed line is
  the putative Fermi surface.}
  \label{nodes}
\end{figure}

We start with the finite temperature ($T\neq 0$) quantum mechanical action for the
interacting $d$-wave quasiparticles,
  \be
  S = S_{\mbox{\scriptsize BCS}} + S_U,
  \ee
where
  \bea
  S_{\mbox{\scriptsize BCS}} = && \swq \w k
  \left[\sum_\s (i\w_n+\xi(\vec k))
    c_\s^\dag(\vec k, \w_n)c_\s(\vec k, \w_n) \right.\nonumber \\
  &&+\D (\vec k) c_\su^\dag(\vec k, \w_n) c_\sd^\dag(-\vec k, -\w_n) + \hc
  \left.\vphantom{\sum_\s}\right],
  \eea
with $\w_n$'s denoting fermionic Matsubara frequencies, and with $\D(\vec{k})$ as the
standard $d$-wave gap. $S_U$ represents the short-range repulsion between electrons,
  \be
  S_U = U\int_0^\b d\t \int d^2\vec{r}
  \left(\sum_\s c_\s^\dag(\vec r,\t)c_\s(\vec r, \t)\right)^2,
  \ee
with $U>0$. Shifting the momenta as $\vec{k}= \vec{K}_{i} +\vec{q}$, $i=I,II$, and
rotating the coordinate frame as in Fig.~\ref{nodes}, we define the eight-component
Dirac field ${\Psi}^\dag=({\Psi_I}^\dag, {\Psi_{II}}^\dag)$ where for each node
$i=I,II$,
  \bea
  {\Psi_i}^\dag(\vec q, \w_n) \equiv \left(
    c_\su^\dag(\vec K_i+\vec q, \w_n),
    c_\sd(-\vec K_i-\vec q, -\w_n), \right. \nonumber \\ \left.
    c_\su^\dag(- \vec K_{i}+\vec q, \w_n),
    c_\sd(\vec K_{i}-\vec q, -\w_n) \right).
  \eea
At $T=0$, with $q_0\equiv\w$, the action may be then written as
  \bea
  S &=&
    \iqe q {\bar\Psi}(q)\G_0\{iq_0+iM(\vec q)\}\Psi(q)
      \nonumber \\
    &+&\,U\itr\t r \left(
        {\bar\Psi}(\vec r, \t) B_U \Psi(\vec r, \t) \right)^2,
  \eea
with $M(\vec q)=\diag(M_I(q_x,q_y),M_{II}(q_y,q_x))$, where $iM_i=\diag(\H_i, \H_{\bar
i})$ is a $4\times4$ matrix defined as
  \be
  \H_i =
  \left( \begin{array}{cc}
    \xi(\vec K_i+\vec q)  & \D(\vec K_i+\vec q)    \\
    \D^*(\vec K_i+\vec q) & -\xi(-\vec K_i-\vec q) \\
  \end{array} \right).
  \ee
Here ${\bar\Psi}={\Psi}^\dag\G_0$, and $\G_0=\diag(\g_0,\g_0)$, with $\g_0 = \s_1\x\I,
\g_1 = -\s_2\x\s_3, \g_2 = -\s_2\x\s_1, \g_3 = -\s_2\x\s_2,$ satisfying the Clifford
algebra $\{\g_\mu,\g_\nu\}=2\delta_{\mu\nu}$. Finally, $B_U=\G_0A_U,$ with
$A_U=\diag(\s_3,\s_3,\s_3,\s_3)$.

One may next expand $iM(\vec q)$ around the nodes of the superconducting order
parameter in powers of $\vec q$. It is convenient to define the following matrices:
$M_1 = -i\s_3\x\s_3,  M_2 = -i\s_3\x\s_1.$ Using the symmetry property of the $d$-wave
gap and the quasiparticle dispersion $\D(-\vec{k})=\D(\vec{k})$ and
$\xi(-\vec{k})=\xi(\vec{k})$, one finds:
  \bea
  &&S_{\mbox{\scriptsize BCS}} = S_0 + S_A + S_{NL}, \\
  &&S_0 = \itr \t r\, {\bar\Psi}(\vec r,\t)\{\G_0\pd\t+
          \G_1\pd x+\G_2\pd y\}{\Psi}(\vec r,\t), \label{S0}\\
  &&S_A = \itr \t r\, {\bar\Psi}(\vec r,\t)\{B_1\pd x+B_2\pd y\}{\Psi}(\vec r,\t),
  \label{SA}
  \eea
where $\G_i=\G_0A_i, B_i=\G_0\d A_i; A_1=\diag(M_1, M_2),$\break $A_2=\diag(M_2, M_1),
\d A_1=\diag(\l_FM_1,\l_\D M_2), \d A_2=\diag(\l_\D M_2,\l_FM_1)$ and $\l_F=v_F-1$,
$\l_\D=v_\D-1$ with $v_F =\pd {q_x} \xi(\vec K_I+\vec q)|_{q=0}, v_\D=\pd {q_y}\D(\vec
K_I+\vec q)|_{q=0}$. $v_F$ and $v_\D$ are the two characteristic velocities of the
linearized spectrum, in units of some fixed velocity $c$, which we set to $c =1$. The
isotropic theory is recovered in the limit $v_F=v_\D$. We will discuss the form of the
$S_{NL}$ that represents the higher order derivative terms shortly. Note that the
matrices $A_i$'s and $\d A_i$'s are independent of $\G_0$.

In the rest of the paper we will consider the linear, isotropic, non-interacting $S_0$
to be the maximally symmetric action. It remains invariant under a global chiral
rotation
 \be
 \Psi \rightarrow e^{i\sum_{i=1}^{16} \theta_i J_i}  \Psi
 \ee
with the generators $J_i, i=1\ldots16,$ forming the $U(4)$ algebra. The explicit form
of the generators is given in~\cite{igor}. The connection to the $QED_3$ is finalized
by coupling the gapless quasiparticles to the fluctuating vortex excitations, which
upon a singular gauge transformation may be represented by an effective gauge field in
$S_0$, by $ \partial_\mu \rightarrow \partial_\mu+i a_\mu$. The charge of the gauge
field is proportional to the vortex condensate~\cite{igor}.  Such a gauge
transformation also turns the $d$-wave quasiparticles into neutral spin-$1/2$
excitations. In the phase coherent, i.e. superconducting state, charge vanishes and
the gauge field is effectively decoupled, leaving the sharp spinon (or quasiparticle)
excitations behind. When the vortices condense, on the other hand, superconductivity
is lost, and simultaneously the spinon "mass" term $\sim m\bar{\Psi}\Psi$ becomes
dynamically generated~\cite{pisarski}. Here, $m=\Sigma(q\rightarrow 0)\neq 0$, with
$\Sigma(q)$ representing the spinon self-energy due to the integraction with the gauge
field. Rewriting the mass term in terms of the electron operators one recognizes that
it simply represents a two-dimensional SDW with the periodicity determined by the wave
vectors $2\vec{K}_{i}$ \cite{igor}.

The SDW order, therefore, corresponds to the particular form of the $\G_0$ matrix we
have chosen. By applying a unitary $U(4)$ transformation on our $\G_0$ one recovers
all other insulating states that are related to the SDW by chiral symmetry. In
particular, transforming $\G_0$ as
  \be
  \G_0 \rightarrow e^{i \frac\pi4 J} \G_0 e^{-i \frac\pi4 J}
  \ee
with $J= \I\otimes \gamma_3$ produces the $d+ip$ insulator: the phase incoherent,
translationally symmetric state with an additional p-wave pairing between spinons.
Similarly, $J=\frac1{\sqrt 2}\s_1\x [(\g_1-\g_2+i\g_0(\g_1+\g_2))]$ yields the $d+is$
insulator, and $J=\frac1{\sqrt 2} \s_2\x (\g_1-\g_2)$ gives the CDW with the
periodicity set by the wavevector $\vec{K}_{I} - \vec{K}_{II}$. Of course, there is a
continuum of other chiral rotations that are possible, and which lead to various
linear combinations of the four fundamental states defined above.

In the rest of the paper we study how the degeneracy between the SDW, CDW, $d+ip$, and
$d+is$ states as defined above is removed by the leading perturbations to $S_0$:
anisotropy, repulsion, and higher order derivatives.

\section{Anisotropy}
We set $U=0$ in $S$ to study first the effect of weak anisotropy $(\lambda_F,
\lambda_{\D} \ll 1)$  on the energy of the degenerate ground states of the chiral
symmetry broken, isotropic $QED_3$. Since the action $S_{\mbox{\scriptsize BCS}}$ is
symmetric under the exchange $v_F \leftrightarrow v_{\D}$, the energies of various
states should be invariant under the same exchange.  Moreover, if $v_F = v_\D$, both
velocities can be rescaled out of the problem by an appropriate choice of $c$. The
energy splittings between the states must therefore be proportional to $(\lambda_F -
\lambda_\D)^2$, to the lowest order. Indeed, defining the three dimensional volume as
$\V=(2\pi)^{-3}\itr \t r$, the first order correction of the energies per volume due
to the velocity anisotropy is
  \bea
  \D E^{(1)}_A &=& \frac1\V\av0{S_1} = \sum_i\tr(B_i\G_i) I_0 \nonumber\\
    &=& 8 m^3(\l_F+\l_\D)I_0,\\
  \eea
independently of the choice of $\G_0$. $I_0$ is a positive, dimensionless integral
  \be
  I_0=\frac13\iq{\Lambda/m}x \frac{x^2}{\s^2(x)+x^2},
  \ee
where $\vec x={\vec q}/m$ and $\s(x) =\Sigma(x) /m$ is the rescaled self energy.
$\Lambda$ denotes the upper cut-off. $\D E^{(1)}_A$ therefore provides only an overall
energy shift, same for all states.

To the second order, however, we find
  \bea
  \D E^{(2)}_A = &&-\frac1{2\V}\left( \av0{S_1^2}-\av0{S_1}^2\right)
               = -\frac12  m^3\sum_{ij}\left[ \tr(B_iB_j)I_{ij} \nonumber \right. \\
                 &&-\tr(B_i\G_\a B_j\G_\b)I_{ij,\a\b} \left. \right], \label{DE21}
  \eea
where
  \bea
  I_{ij}      &=& \iq{\Lambda/m}x \frac{\s^2(x)x_ix_j}{(\s^2(x)+x^2)^2},\\
  I_{ij,\a\b} &=& \iq{\Lambda/m}x \frac{x_ix_jx_\a x_\b}{(\s^2(x)+x^2)^2}.
  \eea
Since the rescaled self-energy $\s(x)$ in the $QED_3$ falls off rapidly for $x\gg 1 $
\cite{pisarski}, we will set $\Lambda=m$ in $I_{ij}$, and then approximate $\s (x) =1$
under the integral. This "constant mass" approximation seems  not to be appropriate
for the integral $I_{ij,\a\b}$, which is without $\sigma(x)$ in the numerator of the
integrand. Luckily, however, $\tr(B_i\G_\a B_j\G_\b)$ is independent of the choice of
$\G_0$ and therefore, the term with $I_{ij,\a\b}$ does not contribute to the energy
differences, but only to the overall shift of the energies. This will be the generic
situation in all further calculations. We will then take $\Lambda=m$ and $\s(x)=1$ in
all the integrals hereafter.

The lowest (second) order effect of the velocity anisotropy is to increase the energy
of the CDW relative to the SDW, d+ip and d+is, which remain degenerate:
  \bea
  &&\D E_{A,\mbox{\scriptsize CDW}}-\D E_{A,\mbox{\scriptsize other}}=4 m^3(v_F-v_\D)^2I>0,\\
  &&I=\frac13 \iq 1x \frac{x^2}{(1+x^2)^2}
  =\frac{10-3\pi}{48\pi^2}.
      \nonumber
  \eea
The energy splitting vanishes when $v_F=v_\D$, as expected.

The fact that $d+ip$ and $d+is$ insulators have the same energy in presence of
velocity anisotropy can be shown to be generally true to any order of the perturbation
theory. To see this, note that $d+ip$ and $d+is$ states are represented by
$\G_{0,d+ip}=\diag(\s_2,-\s_2,\s_2,-\s_2)$, and
$\G_{0,d+is}=\diag(\s_2,\s_2,\s_2,\s_2)$, respectively. When written in the $2\times2$
from, the only difference between the two is in the signs of some terms, which always
may be changed by a unitary transformation. Put differently, the choice of $\G_0$
enters the energy calculation only through the combination $B_i=\G_0\d A_i$. Matrices
$B_i$, on the other hand, have to appear in even numbers in our calculation, as in
Eq.~(\ref{DE21}), otherwise the accompanying integrals will be odd in some momentum
component and vanish. Since $A_i$'s are block-diagonal, the sign of the block-diagonal
elements in $B_i$'s, then, can not matter: $d+ip$ and $d+is$ states remain degenerate
to all orders in weak anisotropy.

\section{Repulsion}
We now set $v_F=v_\D=1 $ to work out the first finite energy contribution of the
short-range repulsion $S_U$ to the degenerate ground states of the isotropic action.
It is found that
  \bea
  && \D E^{(1)}_U = \frac1\V\av0{S_U} = m^4UJ^2\,[(\tr(B_U))^2 -\tr(B_U^2)], \label{DE1U}\\
  && J=\iq 1x \frac 1{1+x^2} = \frac{4-\pi}{8\pi^2}. \nonumber
  \eea
The first term in Eq.~(\ref{DE1U}) vanishes identically for all states; the second
term also vanishes for the CDW, but it increases the energy of the $d+ip$ and the
$d+is$, while decreasing the energy of the SDW
  \bea
  \D E^{(1)}_{U, d+ip}=\D E^{(1)}_{U, d+is}&=& +8 m^4UJ^2,\\
  \D E^{(1)}_{U, \mbox{\scriptsize SDW}}   &=& -8 m^4UJ^2,\\
  \D E^{(1)}_{U, \mbox{\scriptsize CDW}}   &=& 0.
  \eea
Note that $d+ip$ and $d+is$ remain degenerate in presence of the repulsive interaction
as well.

\section{Higher-derivative terms}
We may expand $iM(\vec q)$ beyond the linear terms that led to the anisotropic action,
Eqs.~(\ref{S0}, \ref{SA}).  For instance, defining
  $
  M_\xi = \I\x\s_3,
  M_\D  = \I\x\s_1,
  $
the second-derivative term is given by
  \bea
  S_{NL} = &-&\itr \t r\, {\Psi_I}^\dag \{M_\xi \xi''_0(\pd{}^2)
                                         +M_\D \D''_0(\pd{}^2)\}\Psi_I \nonumber\\
           &+&\{I\rightarrow II, x\leftrightarrow y\} + \ldots,
  \eea
where $\xi''_0(x_1,x_2)=\sum_{ij} x_ix_j\pd{x_i}\pd{x_j}\xi(x_1,x_2)|_{\vec x=0}$ and
similarly for $\D''_0$. Ellipsis indicates higher than second derivative terms. It
seems, then, that we need to specify the dispersion relation and the gap function to
determine the effect of the higher-order derivative terms on the energies of the
degenerate ground states. Interestingly, it turns out that up to the the first
non-vanishing (second) order of perturbation, the qualitative effect of these terms is
to raise the energy of the SDW, lower that of $d+ip$ and $d+is$, and keep the energy
of the CDW unchanged, independently of the functional form of the dispersion and the
gap function.

For definiteness, we present here the results for the tight-binding model on a square
lattice,  for which the dispersion and the $d$-wave gap are given by
  \bea
  \xi(k_a, k_b) &=& -2t(\cos k_aa+\cos k_ba)-\mu \nonumber\\
  \D (k_a, k_b) &=& \D_0(\cos k_aa-\cos k_ba),
  \eea
where $t$  is the nearest-neighbor hopping matrix element, $a$ is the lattice spacing,
$\mu$ is the chemical potential, and $\D_0$ is the amplitude of the $d$-wave
superconducting order parameter. The result is then
  \bea
  \D E^{(2)}_{NL, d+ip} =
  \D E^{(2)}_{NL, d+is}&=& -8 m^3(ma)^2 L,\\
  \D E^{(2)}_{NL, \mbox{\scriptsize SDW}} &=& +8  m^3(ma)^2 L,\\
  \D E^{(2)}_{NL, \mbox{\scriptsize CDW}} &=& 0.
  \eea
Again $L$ is a positive, dimensionless integral, given by
  \bea
  L   &=& \frac{(\cot \tilde{k}_F)^2}{4(ma)^4}\iq 1x \frac 1{(1+x^2)^2}   \nonumber \\
    &\times& \left\{[v_F (\cos (max^+) +\cos (max^-) -2 )]^2 \right. \nonumber \\
      &+&           [v_\D(\cos (max^+) - \cos (max^-))]^2
             \left. \vphantom{[v_F(\cos max^+} \right\},\label{L}
  \eea
where $\tilde{k}_F=\frac1{\sqrt2}k_Fa$ and $x^\pm=\frac1{\sqrt2}(q_x\pm q_y)/m$.
Assuming that near the transition, $ma\ll 1$, to the zeroth order in  $ma$ yields,
  \bea
  L  &=& \frac1{16}(\cot \tilde{k}_F)^2\iq 1x \frac 1{(1+x^2)^2} \nonumber \\
    &\times& \left[ v_F^2(x_1^2+x_2^2)^2+v_\D^2(2x_1x_2)^2\right]
    +O( (ma)^2 ) \nonumber \\
     &=& \frac{\tilde L}{16}(\cot \tilde{k}_F)^2(2v_F^2+v_\D^2)+O( (ma)^2 ),
  \eea
where $\tilde L = (15\pi-46)/180\pi^2.$

\section{Summary.}
Collecting our results together, the shift in energy of a given state may be written
as
  \bea
  \D E = 8 m^3 \left\{\vphantom{ \frac12\theta_A(v_F-v_\D)^2I}\right.
         &&\frac12 \theta_A \,(v_F-v_\D)^2I
        +  \theta_U \,(mU) J^2                              \nonumber\\
        +&&\theta_{NL}\, (ma)^2 L
                 \left. \vphantom{\theta_A \frac12(v_F-v_\D)^2I}\right\},
  \label{expn}
  \eea
where each $\theta_{A,U,NL} = 0$ or $\pm 1$, depending on the change in energy of the
state in consideration. In our constant mass approximation all three integrals $I,J$
and $L$ are mass independent positive constants, so that the Eq.~(\ref{expn})
represents an expansion of the energies of the insulating states in powers of the
dynamically generated mass $m$.

\begin{figure}[t]
  \epsfxsize 50mm
  \epsfbox[5 5 210 220]{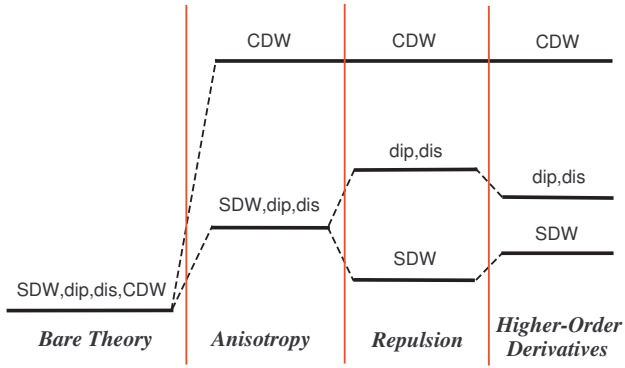}
  \noindent
  \caption[]{Energies of the four types of insulators in the
  unperturbed (bare) action, with the velocity anisotropy, repulsive
  interaction, and the non-linear (higher order derivative) terms.}
\label{fine}
\end{figure}

For weak chiral symmetry breaking, there is a hierarchy of the perturbation terms, in
which each symmetry breaking perturbation assumes a place according to the degree of
its irrelevancy. Velocity anisotropy then, being the marginal perturbation at the bare
level, provides the dominant contribution, while the repulsive interaction, is the
subdominant one. Second order derivatives, although equally irrelevant as the
repulsion, contribute to the energies of the insulating states only to the second
order, and therefore are the least important perturbation. Breaking of the degeneracy
between the insulators due to each perturbation is schematically given on
Fig.~\ref{fine}.

Our conclusion is that the degeneracy of the chiral manifold is broken in favor of the
SDW, which is the lowest energy state when  the weak perturbations to the maximally
chirally symmetric action are taken  into account. The translationally symmetric
$d+ip$ and $d+is$ insulators remain degenerate, as one would expect, since the
translational symmetry remains intact even in presence  of all three perturbations.
Assuming that $d$-wave superconductor-insulator  transition is continuous, or possibly
weakly first order, implies then that the insulating state is the SDW. If the
transition is strongly first  order, chiral mass $m$ increases and it is conceivable
that there could be some level crossings in our  fine structure of the chiral
manifold.

This work was supported by NSERC of Canada and the Research Corporation.

\end{document}